\documentclass[9pt,a4paper,twoside]{rho-class/rho}
\usepackage[english]{babel}
\usepackage{pdfpages}
\setbool{rho-abstract}{true}
\setbool{corres-info}{true}
\makeatletter
\providecommand{\correslanguage}{Corresponding author: }
\providecommand{\emaillanguage}{Email: }
\providecommand{\receivedlanguage}{Received: }
\providecommand{\revisedlanguage}{Revised: }
\providecommand{\acceptedlanguage}{Accepted: }
\providecommand{\publishedlanguage}{Published: }
\providecommand{\@license}{}
\makeatother
\journalname{arXiv}
\title{Potential of large language model-powered nudges for promoting daily water and energy conservation}
\author[1]{Zonghan Li}
\author[2]{Song Tong}
\author[1,3]{Yi Liu}
\author[2]{Kaiping Peng}
\author[1]{Chunyan Wang}
\affil[1]{School of Environment, Tsinghua University, Beijing, China 100084}
\affil[2]{Department of Psychology and Cognitive Sciences, Tsinghua University, Beijing, China 100084}
\affil[3]{State Key Laboratory of Regional Environment and Sustainability, Tsinghua University, Beijing, China 100084}
\leadauthor{Li et al. (2025)}
\corres{Chunyan Wang.}
\email{wangchunyan@tsinghua.edu.cn}

\begin{abstract}
    The increasing amount of pressure related to water and energy shortages has increased the urgency of cultivating individual conservation behaviors. While the concept of nudging, i.e., providing usage-based feedback, has shown promise in encouraging conservation behaviors, its efficacy is often constrained by the lack of targeted and actionable content. This study investigates the impact of the use of large language models (LLMs) to provide tailored conservation suggestions for conservation intentions and their rationale. Through a randomized controlled trial with 1,515 university participants, we compare three virtual nudging scenarios: no nudging, traditional nudging with usage statistics, and LLM-powered nudging with usage statistics and personalized conservation suggestions. The results of statistical analyses and causal forest modeling reveal that nudging led to an increase in conservation intentions among 86.9\%–98.0\% of the participants. LLM-powered nudging achieved a maximum increase of 18.0\% in conservation intentions, surpassing traditional nudging by 88.6\%. Furthermore, structural equation modeling results reveal that exposure to LLM-powered nudges enhances self-efficacy and outcome expectations while diminishing dependence on social norms, thereby increasing intrinsic motivation to conserve. These findings highlight the transformative potential of LLMs in promoting individual water and energy conservation, representing a new frontier in the design of sustainable behavioral interventions and resource management.
\end{abstract}


\begin{document}
\maketitle
    %\thispagestyle{firststyle}
%----------------------------------------------------------

\section{Introduction}

    \rhostart{W}ith urbanization and population growth \cite{RN1}, global domestic water and electricity consumption is projected to increase by 129\% and 120\% by 2050, respectively \cite{RN2,RN3}. Changing individual consumption behaviors is crucial for alleviating supply pressures \cite{RN4} and achieving the Sustainable Development Goals \cite{RN5}. Currently, the promotion instruments used by governments, including economical \cite{RN6,RN7}, educational \cite{RN8,RN9}, and technical approaches \cite{RN10,RN11}, often prove to be costly or inefficient. The concept of nudging, which is a transferable and widely used tool for altering individual behaviors, offers a promising alternative. By providing indirect information or modifying the choice environment \cite{RN12}, nudges have shown to produce small to moderate effects \cite{RN13,RN14} across various domains \cite{RN15,RN16,RN17,RN18}. There is ample evidence suggesting that nudging can reduce water and energy consumption by 1\%–15\% in residential, hotel, and university contexts \cite{RN19,RN20,RN21}. However, the effectiveness of nudges varies significantly depending on the content type, and the pathways by which they influence the decision process remain unclear.

    Nudges for individual water and energy conservation or conservation intentions (the willingness to exert effort to perform a behavior rather than actual behaviors \cite{RN22,RN23}) are typically conducted by providing usage-related feedback \cite{RN19,RN20,RN21}. The main content categories include (1) current and historical usage that can increase individuals’ awareness of usage patterns \cite{RN24}; (2) social comparisons that can encourage conservation behaviors by activating social norms through comparisons with similar individuals or groups \cite{RN25,RN26}; and (3) conservation tips that can facilitate specific conservation behaviors, which usually consist of standardized general advice for all individuals \cite{RN27,RN28}. Current nudges largely rely on usage statistics (i.e., categories 1 and 2), especially social comparisons. Only a few offer conservation tips and rarely customize these tips on the basis of usage statistics \cite{RN27,RN28}. However, given the heterogeneity of residents’ usage patterns, customizing conservation tips through usage statistics can help target behaviors with the highest conservation potential, thereby enhancing the effectiveness of nudging. Despite the wide encouragement of this customization \cite{RN29,RN30}, recent work has continued to focus on more detailed usage statistics \cite{RN31} or eye-catching delivery modes \cite{RN32,RN33,RN34} rather than intelligent and tailored nudging content.

    Owing to their strong capabilities in data analysis \cite{RN35}, text generation \cite{RN36}, and interaction with individuals \cite{RN37}, large language models (LLMs) can analyze individual water and energy usage patterns to provide more intelligent and personalized content accordingly. LLMs have been preliminarily applied in various fields (e.g., finance \cite{RN38} and education \cite{RN39}) to provide knowledge or behavioral interventions, such as providing LLM-generated images to improve support for policies \cite{RN40}. When providing nudges regarding water and energy conservation, LLMs can help analyze behavioral patterns, select the most effective and actionable conservation tips, and provide detailed individualized estimates of conservation potential, thereby potentially enhancing the effectiveness of nudges. However, the effectiveness of LLM-powered nudges remains unclear.

    With respect to explaining the decision processes through which nudges operate, previous studies have identified individual water and energy conservation as both cognitively and socially influenced, with a wide range of psychological antecedents playing vital roles. For example, cognitive factors, including personal norms \cite{RN41, RN42}, outcome expectations \cite{RN43, RN44}, and abilities or perceived difficulties \cite{RN45, RN46}, could intrinsically either motivate or hinder conservation, while social influences such as neighborhood perceptions \cite{RN47, RN48} help reinforce peer comparisons, and attitudes \cite{RN49, RN50} help link cognitive factors to conservation and shape individuals’ conservation decisions. On the basis of this knowledge, quantifying how individuals make decisions under LLM-powered nudges and how these nudges affect these antecedents can provide insights into the role of LLMs and guide the optimization of future interventions.

    This study aims to measure the changes in the conservation intentions of individuals who receive different types of nudges and explore how LLM-powered nudges function with lower time and monetary costs. We use a university campus as a case study because of the global attention to campus sustainability \cite{RN51,RN52,RN53,RN54}, high per capita water and energy use in campus buildings (5–10 times higher than that of residential buildings \cite{RN55,RN56}), and the generally low level of student engagement in conservation behaviors \cite{RN55,RN57}. These findings can offer new insights and cutting-edge tools for resource management and determine the potential role of LLMs in altering individual conservation behaviors. As university campuses represent significant urban populations \cite{RN58} and are often used as models or testing grounds for urban policies \cite{RN59,RN60}, considering the transferability and cost-effectiveness of LLM-powered nudges, the implications of this study may extend beyond campuses to broader urban contexts.
    
\begin{figure*}
  \centering
  \includegraphics[width=1\textwidth]{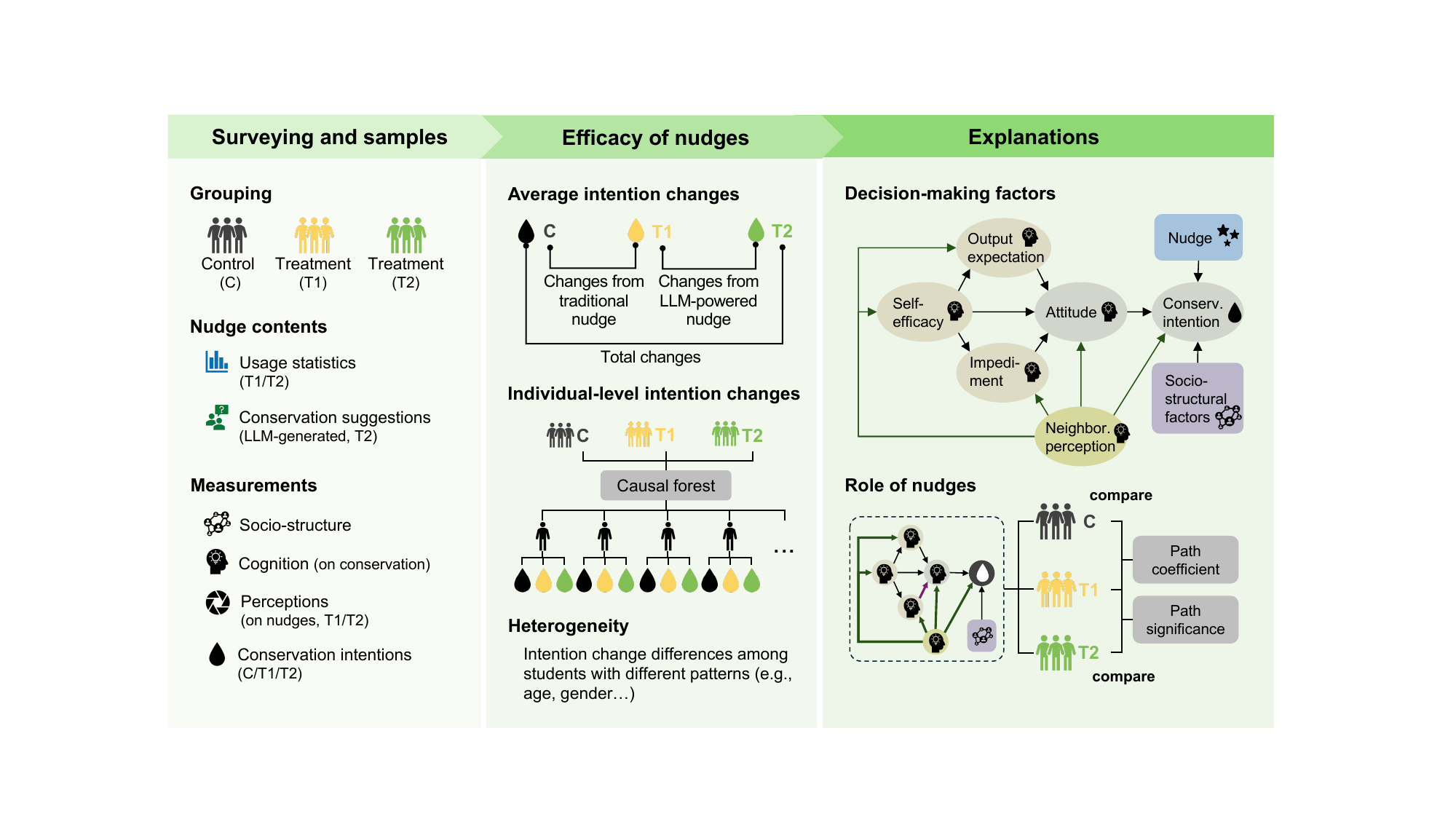}
  \caption{Roadmap of this study.}
  \label{fig1}
\end{figure*}

\section{Results}
    To examine the effectiveness of the proposed nudging approach with LLMs, in this study, we took intentions on water and energy conservation (abbr. conservation intention) as an example and followed the roadmap illustrated in Figure \ref{fig1}. Specifically, we first measured and compared conservation intentions using a survey experiment conducted on a university campus. We then identified the effectiveness of traditional and LLM-powered nudges through statistical tests and causal forests. Finally, we delineated the LLMs’ functions in changing conservation intentions through a series of structural equation models.

    In the experiment, virtual nudge scenarios (i.e., demonstrations of nudges) with different combinations of nudge contents were delivered to three randomly assigned groups of respondents. The experiment was conducted through a questionnaire survey that received 1,515 responses (1,416 valid). Before they participated in the experiment, the respondents were asked to report various characteristics in the questionnaire, including personal information (i.e., socio-structural factors) and psychological factors related to water and energy conservation (i.e., cognition factors). Only then did they receive demonstrations of nudges and, on the basis of these demonstrations, report their conservation intentions and attitudes toward the nudges. The three groups consisted of the Control Group (C), which was not provided with any energy or water-related content; Treatment Group 1 (T1), which received traditional nudge content featuring current and historical usage data and social comparisons; and Treatment Group 2 (T2), which received LLM-powered nudges that incorporated conservation tips and potential estimations in addition to traditional nudge contents (see details in the Methods section).

\subsection{Descriptive statistics}

    Prior to the experiment, we measured socio-structural factors (including demographics, daily energy behaviors, and living metrics), historical water and energy conservation habits, and students’ perceptions of water and energy conservation (i.e., thoughts and beliefs that influence conservation). Additionally, we surveyed different groups’ attitudes toward nudges to determine students’ acceptance and perceived effectiveness of nudges.

    Among the participants, 36.6\% were female (standard deviation calculated with normalized data, SDN = 0.48). Undergraduates comprised 55.7\% of the sample, whereas master’s and doctoral candidates represented 16.9\% and 27.4\%, respectively (SDN = 0.43). A majority (70.1\%) of the participants had prior experience paying electricity bills while living in their dormitories (SDN = 0.46). Additionally, 85.5\% of the participants reported commuting on campus by cycling or walking instead of using e-bikes (SDN = 0.35). The average monthly living cost for the students was 2,116.9 CNY (SDN = 0.23), and the students reported spending an average of 11.3 hours daily in their dormitories (SDN = 0.21).

\begin{figure*}
  \centering
  \includegraphics[width=1\textwidth]{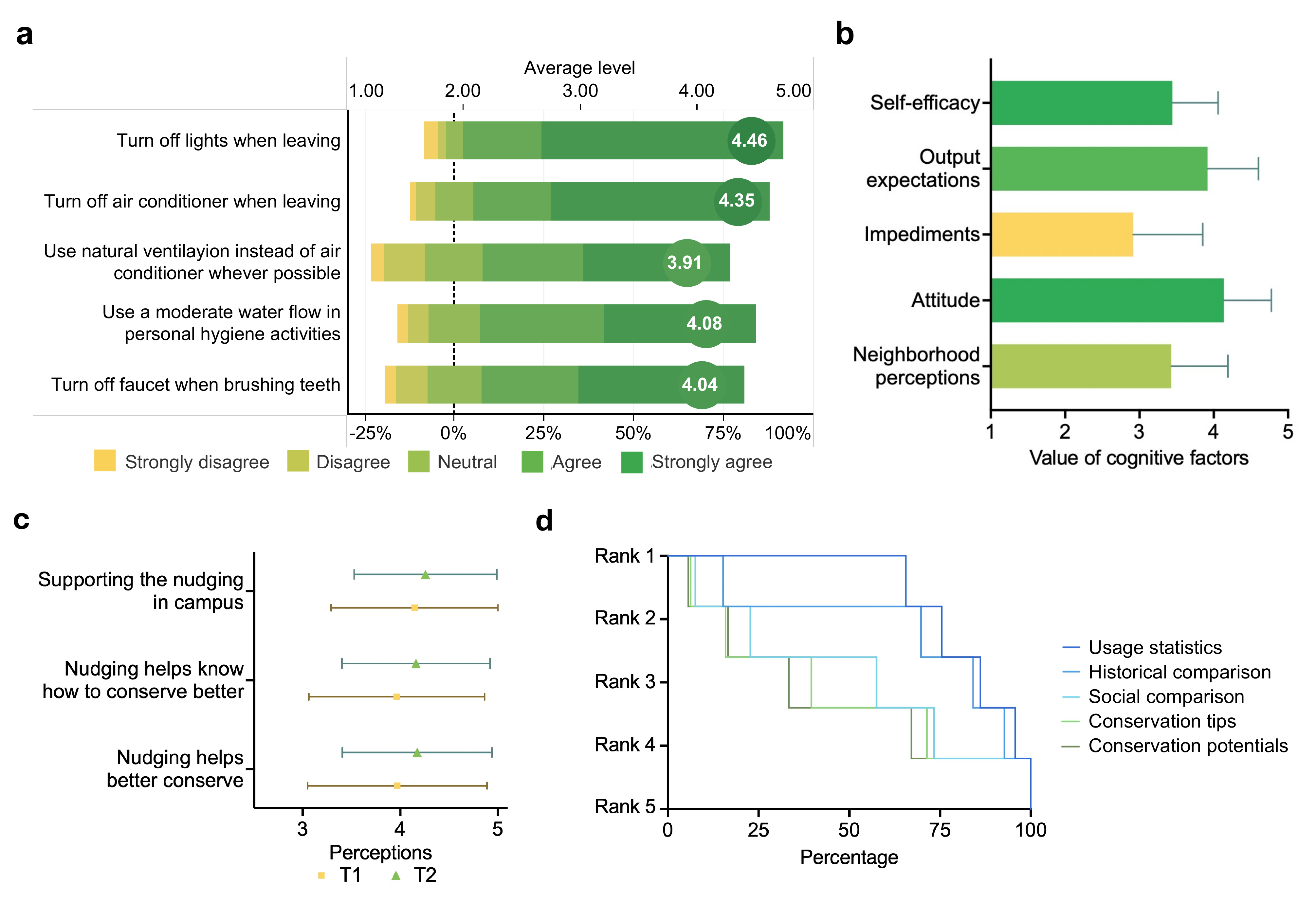}
  \caption{\textbf{Descriptive statistics of the questionnaire survey.} \textbf{a.} Historical behaviors related to water and energy conservation, measured using a 5-point Likert scale. The upper axis shows the average level of these behaviors, whereas the lower axis displays the percentage of respondents selecting each option. \textbf{b.} Cognition factors related to water and energy conservation, with error bars representing the 95\% confidence intervals. \textbf{c.} Attitudes toward nudges, with error bars representing the 95\% confidence interval. \textbf{d.} The rated effectiveness of different nudge contents.}
  \label{fig2}
\end{figure*}

    With respect to the most significant energy and water consumption within dormitories, participants exhibited a relatively high level of conservation behavior (Figure \ref{fig2}a), with energy conservation showing slightly greater adherence than water conservation. Specifically, 89.2\% of the participants reported turning their lights out when leaving their dormitories (average score = 4.46 out of 5; SDN = 0.24). Additionally, 82.8\% reported turning off their air conditioners (average score = 4.35; SDN = 0.24), and 68.8\% opted to use natural ventilation instead of air conditioning (average score = 3.91; SDN = 0.29), which also contributed to energy savings. In contrast, water conservation behaviors were less common and more varied, with average scores of 4.08 (SDN = 0.26) for moderate water flow and 4.04 (SD = 0.28) for turning off faucets when not in use.

    Most of the cognitive factors related to water and energy conservation were favorably rated (Figure \ref{fig2}b). Attitudes were found to be the most positive, with an average score of 4.13 out of 5. The output expectations followed closely, with an average score of 3.92, reflecting a high level of anticipated benefits of conservation behaviors. The average scores of neighborhood perceptions and self-efficacy were 3.43 and 3.44, respectively, which indicated moderate levels of both perceived conservation in students’ social networks and confidence in their conservation ability. Impediments (reverse scored) scored lowest at 2.92, suggesting a relatively low level of perceived difficulty in conservation.

    The participants expressed positive attitudes toward the nudges, with an average score of 4.11 out of 5, as shown in Figure \ref{fig2}c. Notably, the participants in T2 had an average attitude score of 4.20, which was 4.5\% higher than that at T1 (4.02), indicating that LLM-powered nudges were considered more acceptable and effective. For the rating for nudging content on the basis of the results of the T2 samples, as illustrated in Figure \ref{fig2}d, usage statistics were considered the most important information, with 65.6\% of the participants ranking this content first. Historical comparisons and social comparisons were the second and third types of recognized information, respectively. However, the conservation suggestions generated by the LLMs were considered the least important, with average rankings of 3.67 and 3.77.

\subsection{LLM-powered nudging enhanced conservation intentions}

\begin{figure*}
  \centering
  \includegraphics[width=0.85\textwidth]{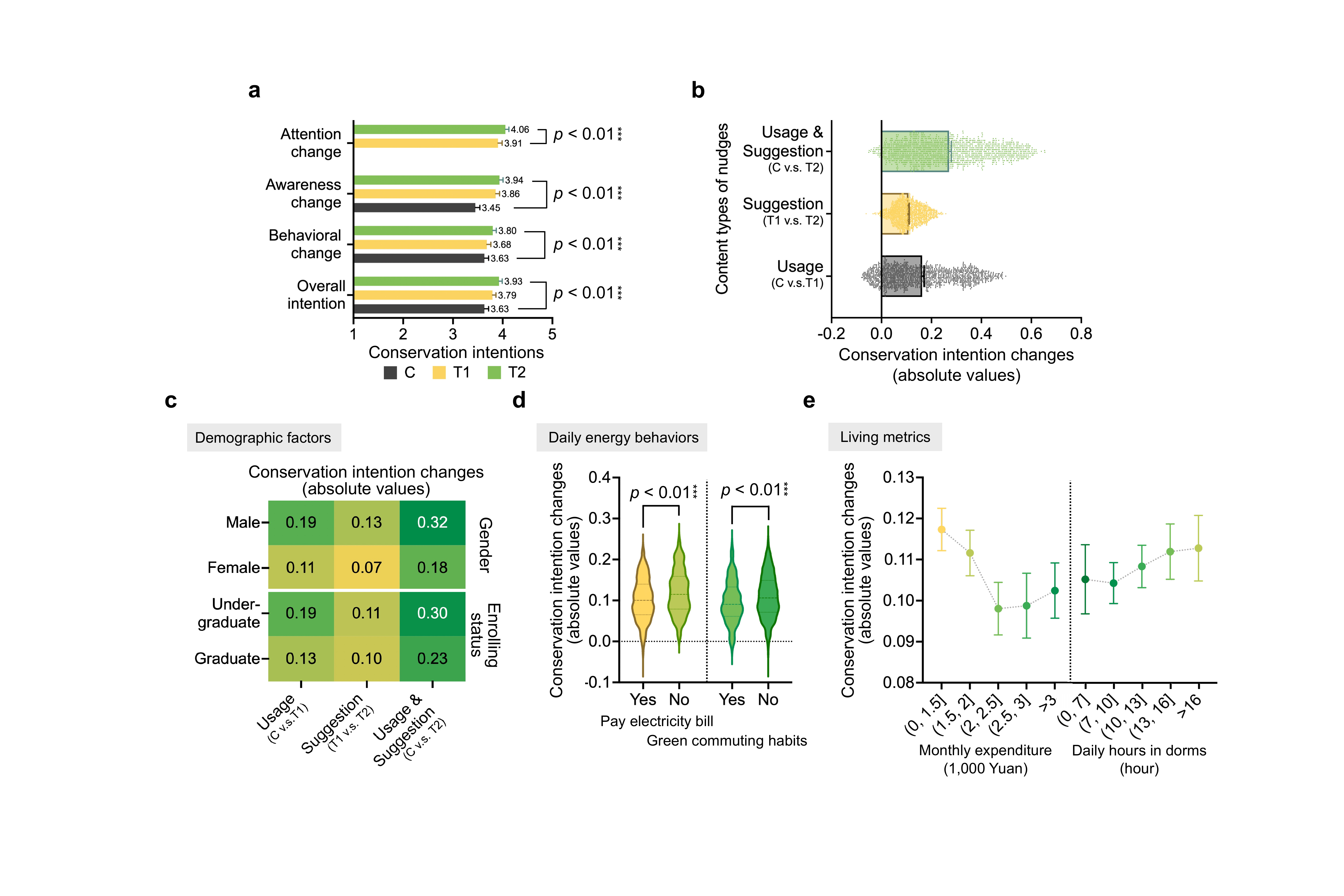}
  \caption{\textbf{Increases in conservation intentions caused by nudges and their heterogeneity. a.} Average increases according to the statistical comparison among the three groups, with error bars representing the 95\% confidence intervals. \textbf{b.} Individual-level conservation intention changes and their distributions calculated by the multi-arm causal forest. The error bars represent the 95\% confidence intervals. \textbf{c.} Ranges of conservation intentions increase in different demographic groups. \textbf{d.} Ranges of conservation intentions increase in student groups with different daily energy behaviors. \textbf{e.} Ranges of conservation intentions increase in student groups with different living metrics, with error bars representing the 95\% confidence interval.}
  \label{fig3}
\end{figure*}

    The changes in conservation intentions caused by different types of nudges and the heterogeneity of these changes are illustrated in Figure \ref{fig3}. We used statistical comparisons among the three groups to reveal the average changes and causal forests to infer changes at the individual level. Conservation intentions were measured through three aspects: attention (reading the nudge contents), awareness (paying attention to daily usage), and behavior (implementing conservation actions). Details are provided in the Methods section.

    As shown in Figure \ref{fig3}a, the participants generally exhibited high levels of conservation intentions, with average scores for C, T1 and T2 being 3.63, 3.79, and 3.93, respectively, out of a total of 5. The mean value and distribution patterns of conservation intentions among the three groups were statistically significant (\textit{p} < 0.01), indicating the enhancement caused by nudging. Specifically, the increase of 8.3\% between C and T2 illustrates that LLM-powered nudges indeed enhanced conservation intentions. With respect to different types of nudges, traditional nudges increased conservation intentions by 4.4\% (from 3.63 to 3.79, C vs. T1, \textit{p} = 0.05), whereas LLM-powered nudges amplified the effectiveness of traditional nudges by 88.6\%, further increasing conservation intentions by 3.9\% (from 3.79 to 3.93, T1 vs. T2, p = 0.05). These nudge types targeted different aspects of conservation intention; traditional nudges primarily influenced awareness changes (from 3.45 to 3.86), whereas LLM-powered nudges affected behavioral changes (from 3.68 to 3.80). Interestingly, these increases may have been elevated by LLM-powered nudges without the participants’ conscious awareness given that the participants in T2 perceived LLM-generated conservation suggestions as the least important components (Figure \ref{fig2}d). This aligns with the finding of previous research in other fields, which highlights that nudging can be effective even at the unconscious level \cite{RN61,RN62, RN63}. 

    We further employed a multi-arm casual forest to gain a deeper understanding by estimating the potential changes in the intentions of each participant for each type of nudge (i.e., individual treatment effects). The causal forest algorithm \cite{RN64}, which is an extension of random forests \cite{RN65}, is widely used to estimate the individual treatment effects of interventions in many domains, such as household energy use \cite{RN66}, education \cite{RN64}, and agriculture \cite{RN67}. As illustrated in Figure \ref{fig3}b, the multi-arm causal forest yielded similar conservation intention changes but provided more detailed individual-level estimations. For LLM-powered nudges, the individual-level average estimated conservation intention change amounted to an increase of 0.27, corresponding to a 7.7\% increase compared with the average for C. The maximum increase observed among the participants was 0.65, representing an 18.0\% increase. Conservation intention changes basically followed a normal distribution, with 92.4\% of the participants showing improved conservation intentions (86.9\% for T1 vs. 98.0\% for T2), indicating that nudging can effectively promote positive outcomes among most university students. Moreover, each type of nudge content also contributed differently to the increases; for usage-related content, the type increased the conservation intentions by an average of 0.16 (4.5\%) and by as much as 0.50 (13.7\%) at most, whereas for conservation suggestions, based on the foundation set by usage-related content, the type further raised conservation intentions by an average of 0.11 (3.0\%) and a maximum of 0.26 (7.1\%).

    The findings from both the statistical comparisons and casual forests indicate that nudging can increase the conservation intention of the participants and that LLM-powered nudging can further “catalyze” the enhancement through traditional nudging. Furthermore, we utilized the individual-level changes in conservation intentions calculated through the multi-arm causal forest to explore the characteristics of socio-structural groups that exhibit various increases in conservation intentions. The results that are applicable to various types of nudging content are summarized in Figure \ref{fig3}c, whereas Figure \ref{fig3}d presents results applicable only to the conservation suggestions in LLM-powered nudges.

    Several differential patterns of demographic factors emerged (Figure \ref{fig3}c). First, the increases were found to be more pronounced in males. Specifically, regardless of the content type of nudges, the average conservation intention change of males was significantly greater than that of females (e.g., 0.32 vs. 0.18 for C vs. T2, \textit{p} < 0.01). Second, the effects varied across different educational levels, with nudging having a greater impact on undergraduates than on graduate students, including both master’s and doctoral students (e.g., 0.30 vs. 0.23 for C vs. T2, \textit{p} < 0.01).

    Daily energy behaviors and living metrics also exhibited heterogeneity, primarily generated by conservation suggestions for LLM-powered nudges (Figure \ref{fig3}d and Figure \ref{fig3}e). First, participants who reported being responsible for paying electricity bills (i.e., are more aware of their electricity usage) were found to be less responsive to the suggestions than those who were less aware (0.10 vs. 0.12, \textit{p} < 0.01). Second, the effectiveness of nudging was closely related to other environmental behaviors. Participants who reported usually choosing to travel by bicycle rather than e-bikes were more likely to exhibit greater increases in conservation intention (0.11 vs. 0,10, \textit{p} < 0.01). Third, conservation intention changes were found to be influenced by monthly expenditures. Among those with high increases in conservation intentions, lower daily expenditures were found to be associated with greater conservation intention changes (e.g., 0.12 for < CNY 1,500 vs. 0.10 for > CNY 3,000, \textit{p} < 0.01), indicating that nudges are more effective for price-sensitive participants. Finally, changes in conservation intention were found to be related to the time spent in the dormitory. The participants whose conservation intentions increased reported spending slightly longer times in their dormitory (e.g., 0.10 for 7–10 hours vs. 0.11 for > 16 hours, \textit{p} = 0.06). Although the difference appears small, after excluding average sleep time, the additional time is logical, as longer exposure to the shared environment with other members can make behaviors more susceptible to the influence of roommates.

\subsection{Role of LLM-powered nudges in forming conservation intentions}

\begin{figure*}
  \centering
  \includegraphics[width=0.9\textwidth]{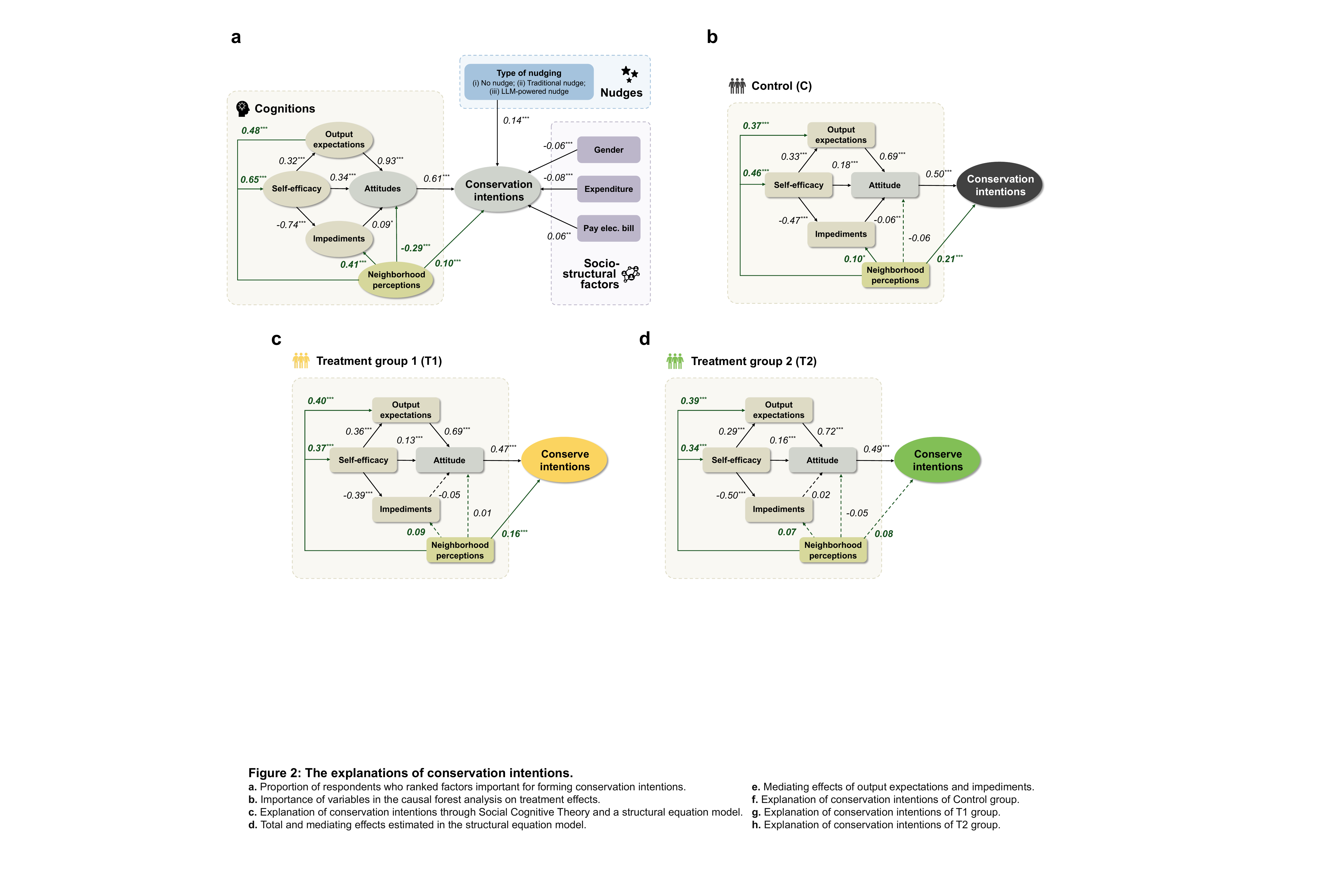}
  \caption{\textbf{Decision-making related to conservation intentions. a.} The full model of the decision-making of conservation intentions under nudges. \textbf{b.} The decision-making of conservation intentions among samples with no nudge (C).\textbf{ c. }The decision-making of conservation intentions among samples with traditional nudges (T1). \textbf{d.} Decision-making of conservation intentions among samples with LLM-powered nudges (T2).}
  \label{fig4}
\end{figure*}

    We used structural equation modeling (as shown in Figure \ref{fig4}) to simulate the formation process of individual conservation intentions. As stated in the main text, conservation intentions are influenced mainly by attitudes, a series of self-cognition factors, neighborhood perceptions, and socio-structural factors. We mainly referred to Bandura’s theory of social cognition \cite{RN68} and Ajzen and Fishbein’s theory of reasoned action \cite{RN69} to establish the models.

    First, we established a \textit{full model} to determine the formation mechanism of conservation. Owing to limitations in statistical modeling caused by statistical power constraints, only the factors within each type that had greater variance after 0–1 normalization and thus could provide more information \cite{RN70, RN71, RN72} were selected for modeling. The model fit the data well ($\chi^2$(df) = 5.78, RMSEA = 0.06, CFI = 0.94, SRMR = 0.04). The \textit{type of nudging} (i.e., no nudges, traditional nudges, or LLM-powered nudges) positively predicted conservation intentions ($\beta$ = 0.14, \textit{p} < 0.01), revealing that with more comprehensive nudge content, students are more likely to have greater conservation intentions. In addition, conservation intention was found to be positively affected by \textit{attitude}, which is the key conclusion based on the theory of reasoned action. Among \textit{self-cognition factors}, self-efficacy was found to be a primary factor that not only directly positively affected attitudes ($\beta$ = 0.34, \textit{p} < 0.01) and positively impacted output expectations but also negatively impacted impediments. In addition to self-efficacy, attitudes were also significantly positively influenced by output expectations ($\beta$ = 0.93, \textit{p} < 0.01), whereas the impact of impediments on attitudes was comparatively minor and not considered a major factor affecting attitudes under nudges due to limited significance ($\beta$ = 0.09, \textit{p} < 0.05). \textit{Neighborhood perceptions} positively affected conservation intentions ($\beta$ = 0.10, \textit{p} < 0.01) and all cognition factors ($\beta$ = 0.46 on average, all \textit{p} < 0.01), indicating the importance of social networks in the development of conservation intentions.

    The differences in the same paths under various nudge types were compared to further reveal the effectiveness of nudging in shaping conservation intentions. Three groups of structural equation models were established for C, T1, and T2, as shown in Figure \ref{fig4}b, Figure \ref{fig4}c, and Figure \ref{fig4}d, respectively; they fit the data well ($\chi^2$(df) = 4.72, RMSEA = 0.08, CFI = 0.93, SRMR = 0.07). The socio-structural parts are not presented in the figures to make them clearer; their details can be found in Supplementary Table S7.

    Differences in both social factors and self-cognition were identified. The importance of neighborhood perceptions as a manifestation of social networks was examined through the full model. However, with the contents of nudges becoming more tailored and detailed, neighborhood perceptions seem to have lost their impact on both conservation intentions and other cognitive factors. The path coefficient from neighborhood perceptions to conservation intentions was 0.21 (\textit{p} < 0.01) for C, 0.16 (\textit{p} < 0.01) for T1, and 0.08 (\textit{p} > 0.10) for T2, while the average coefficient of neighborhood perceptions to other cognitive factors (paths marked in dark green) was 0.31 for C (\textit{p} < 0.01), 0.29 for T1 (\textit{p} < 0.01), and 0.20 for T2 (\textit{p} < 0.01). The decreasing values and statistical significance of path coefficients indicate that nudges, especially LLM-powered ones, would potentially weaken or even replace the role of social networks in the formation of conservation intentions.

    LLM-powered nudges were found to change the impact of self-cognition in the decision-making process by weakening the negative impact of impediments and enhancing the power of both self-efficacy and output expectations. As a path with relatively a low level of significance in the full model, impediments were found to have a negative effect on attitudes in C ($\beta$ = -0.06, \textit{p} = 0.05); however, this effect became nonsignificant at T1 (\textit{p} = 0.14) and T2 (\textit{p} = 0.60) under nudges, revealing that LLM-powered nudges could further decrease the adverse effects of impediments when forming conservation intentions. For both self-efficacy and output expectations, the coefficient of their impact on attitudes increased by 28.6\% (from 0.13 to 0.16) and 3.9\% (from 0.69 to 0.71), respectively, indicating that LLM-powered nudges can further increase the effectiveness of self-cognitive factors, i.e., intrinsic motivations, for conservation.

\section{Discussion}

    This study builds on existing nudge content by examining whether the use of emerging LLMs can provide more intelligent and tailored nudge content to achieve greater increases in conservation intentions. LLM-powered nudges were found to enhance the conservation intentions of 98.0\% of individuals by an average of 8.3\% and up to 18.0\% by altering the influence of self-cognition and social factors. Notably, the savings resulting from increased conservation intentions are not nontrivial. For example, if students could fully implement these conservation behaviors, then the case study university could save approximately 3.36 million kWh of electricity and 0.11 million cubic meters of water per year. Extending this approach to all higher education institutions in China could lead to annual reductions in electricity and water consumption of approximately 1.14 billion kWh and 0.04 billion cubic meters, respectively. These savings are equivalent to the annual electricity consumption of approximately 1.20 million residents and the water usage of approximately 0.86 million individuals (calculations detailed in Supplementary Note S7). Notably, these savings could be achieved at minimal costs.

    Our results have three main implications. First, the value of using LLMs in water and energy conservation nudges is highlighted. Traditional nudges increased conservation intentions by an average of 4.4\% (up to 13.7\%), whereas LLM-powered nudges enhanced the effectiveness of traditional nudges by 88.6\% (average increase: 8.3\%, maximum 18.0\%), illustrating the significant potential of LLM-powered nudges in promoting water and energy conservation. Generally, nudges alter individual decisions by triggering System 1 (i.e., inducing intuitive responses that lead to different choices), engaging System 2 (i.e., prompting individuals to engage in more reflective cognitive processes through brief pauses), or bypassing both systems (i.e., removing individuals from certain aspects of the decision-making process) \cite{RN73, RN74, RN75}. For water and energy conservation, traditional nudges primarily inform individuals about their consumption status and activate social norms, thereby altering System 2. LLM-powered nudges motivate and guide individuals toward specific sustainable behaviors, thereby influencing System 1. Simultaneously, altering both systems can more effectively promote desired outcomes, i.e., increase conservation intentions, and thus increase the amount of saved water and energy.

    Second, the results also contribute to a deeper understanding of the effects of nudges. Herein, nudges motivated a substantial portion of the participants to increase their conservation intentions to some degree rather than merely influencing a small subset of people. Moreover, LLM-powered nudges had a greater overall impact on the study population than traditional nudges did (98.0\% vs. 86.9\%). However, consistent with other studies on nudging, a small fraction of individuals was resistant, reacting in the opposite direction to the encouragement \cite{RN15}. Additionally, our findings suggest that students with certain social characteristics (male, undergraduate), daily behaviors (spending more time in dormitories, tighter daily expenditures, engaging in other pro-environmental behaviors), and cognitive and perceptual traits (lower conservation cognition and less positive attitudes toward nudges) are more responsive to nudges. This information can be utilized to improve the targeted training of LLMs and to set intervention goals for future real-world interventions.

    Third, the study sheds light on the psychological mechanisms underlying participants’ responses to nudges and the role of LLMs behind the responses. The formation of conservation intentions of the participants is influenced by self-cognition, social cognition, the type of information conveyed by the nudge, and other social-structural characteristics. Traditional nudges heavily rely on social norms, where individuals are often influenced by peer behaviors or societal expectations \cite{RN76, RN77}, as reflected by the decreasing role of neighborhood perceptions across the treatment groups. In contrast, the personalized nature of LLM-powered nudges reduces reliance on external social cues, allowing students to make more autonomous decisions. By addressing perceived impediments and emphasizing the personal relevance and feasibility of conservation actions, LLM-generated suggestions offer clearer, actionable steps that are tailored to individual circumstances. This improved suggestion not only impacts the three critical drivers of sustainable behavior—thereby enhancing self-efficacy and outcome expectations while reducing perceived impediments—but also strengthens intrinsic motivation, i.e., conservation intentions. As Damgaard and Nielsen \cite{RN78} noted, traditional nudges, while effective, often lead to short-term behavioral shifts without deeply influencing intrinsic motivation. By altering the impact of self-cognitive factors, LLM-powered nudges have the potential to stimulate more enduring intrinsic motivation, which in turn drives sustainable and long-term behavior change. This approach applies not only to water and energy conservation but also to broader contexts in education and social behaviors. Through dynamic and personalized feedback, LLM-powered nudges can more effectively promote critical cognitive drivers, resulting in lasting and meaningful behavior change. This shift from socially driven to cognitively driven decision-making could mark a new frontier in the design of sustainable behavioral interventions.

    Some limitations remain in this study. As a preliminary experiment on intentions, although we strictly controlled the randomness of group assignments and employed stringent data quality control standards, the provided virtual scenarios and LLM-generated demonstrations may still present a gap when compared with real behaviors. Second, this study emphasized only the impact of intervention type on conservation intentions, without considering the effects of other variations (such as feedback frequency, feedback media, etc.). Finally, we were unable to examine whether water and energy conservation nudges had spillover effects on other pro-environmental behaviors. Such spillover could be a part of the nudge effects that may have been overlooked in the current study.

\section{Methods}

    The study complied with all ethical regulations and was approved by the Tsinghua University Science and Technology Ethics Committee. Informed consent was obtained from all participants.

\subsection{Experiment and surveying}

\begin{figure*}
  \centering
  \includegraphics[width=1\textwidth]{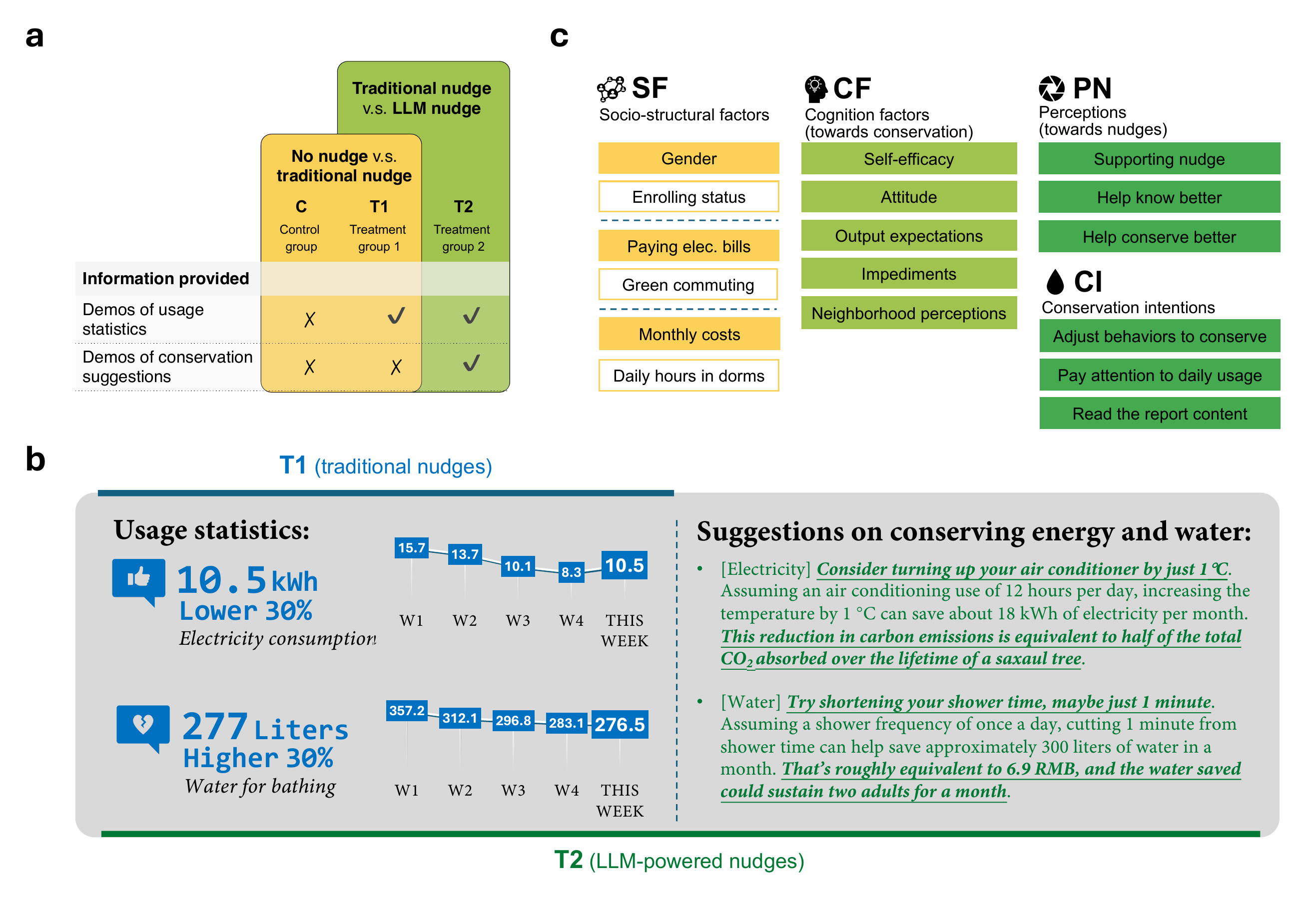}
  \caption{\textbf{Measurements and experimental design. a.} Groupings and included demonstrations of nudge content used to measure conservation intentions. \textbf{b.} An example of a demonstration of a nudge. \textbf{c.} Measurements used in the questionnaire survey.}
  \label{fig5}
\end{figure*}

    The survey experiment was carried out in Questionnaire Star (https://www.wjx.cn), which is a questionnaire survey platform used in mainland China. The participants were first asked to submit their personal information (i.e., socio-structural factors, 6 questions in total) and psychological factors related to water and energy conservation (i.e., cognition factors, 5 groups of questions in total) and then to report their conservation intentions (3 questions). As shown in Figure \ref{fig5}a, for the conservation intention portion of the study, three groups were designed, and participants were randomly assigned to one of the three groups, namely, the Control Group (C), which received no nudge information; Treatment Group 1 (T1), which received demonstrations of usage statistics (including current usage, historical usage and social comparisons, as illustrated in the left part of Figure \ref{fig5}b); and Treatment Group 2 (T2), which received demonstrations of both usage statistics and LLM-generated conservation suggestions (including conservation tips and corresponding conservation potentials, as displayed in the right part of Figure \ref{fig5}b). The participants were each randomly assigned to a group by the survey platform when they entered the survey and were unaware of the grouping. This approach meets the standards of random assignment and concealment in experiments, thereby minimizing the level of possible bias \cite{RN79}. Additionally, differences among the three groups were tested to ensure randomness (detailed in Supplementary Note S1 and Supplementary Figure S1).

    The demonstrations of the nudge scenarios were generated by a large language model-based nudging platform developed by the authors. The platform helped analyze the behavioral patterns of students and provide effective and actionable conservation suggestions to students through answers given to knowledge-based question and content recommendations, including estimates of conservation potential. Although real usage statistics were not used, participants in both T1 and T2 received detailed explanations and examples of the types of analysis data. Additionally, participants in T2 were provided with information on how the suggestions were generated by a large language model and customized on the basis of individual characteristics, along with the demonstrations. %(detailed in Supplementary Note S1).

    Conservation intentions were collected through the question “\textit{What would be your intention to engage in the following behaviors?}” after viewing the provided statement “\textit{Following the current water and energy conservation promotion policies at the university…}” for C or “\textit{If provided with the following information generated by large language models according to your usage statistics each week…}” for T1 and T2, along with the demonstrations of nudge contents. The role of traditional nudges could be identified by comparing the results of C and T1, and the effectiveness of LLM-powered nudges could be highlighted by comparing the results of T1 and T2.

    The participants of this study included 1,515 students at Tsinghua University, Beijing, China, who lived in dormitories provided by the university. The participants were recruited in June 2024 using a combination of purposive and snowball sampling techniques, both of which have been widely used in other surveys conducted at universities \cite{RN80}. Posters and QR codes advertising the questionnaire were posted both on dormitory bulletin boards and in student WeChat groups by student community officials in all the undergraduate, master’s, and doctoral dormitory buildings. Each participant received a red packet with a random amount of e-cash equivalent that varied from \$1.5 to \$2.5 after the questionnaire was deemed valid.

    We excluded participants who either did not follow the instructions (determined by checking whether mandatory selection questions were answered as instructed), provided inconsistent responses, or submitted the questionnaire in an unreasonably short answer time. The remaining sample size was 1,416 (valid response rate of 93.5\%).

\subsection{Measurements}

    The questionnaire survey assessed socio-structural factors, cognition factors related to conservation, attitudes pertaining to nudges, conservation intentions, and other variables, as illustrated in Figure \ref{fig5}c.

    \textbf{(1) Socio-structural factors.} We measured three types of socio-structural factors. The first type consisted of demographic factors, including \textit{gender} (binary) and \textit{enrollment status} (1-undergraduate, 2-master’s, 3-doctoral). The second type consisted of daily energy behaviors, including experiences \textit{paying electricity bills} (binary, 1-yes, 0-no) and habits of \textit{green commuting} (binary, 1-yes/by bike/on foot, 0-no/by other). In each dormitory at Tsinghua University, 2–4 students share one electricity meter. There is a fixed quota of free electricity each semester. Once the free electricity is depleted, one dormitory resident must be responsible for paying the electricity bill through the school’s payment system. Since the electricity bill shows the dormitory’s power usage, we used this experience to gauge whether students were aware of their dormitory’s electricity consumption. The third type consisted of living metrics, including \textit{monthly} expenditures and \textit{daily hours in dorms} (both continuous).

    \textbf{(2) Cognition factors.} Five measurement scales were developed to measure cognition factors (i.e., self-efficacy, output expectations, perceived impediments/facilitators, attitudes, and neighborhood perceptions). Unless otherwise noted, participants were instructed to rate their agreement with each statement on a Likert scale ranging from 1 (‘strongly disagree’) to 5 (‘strongly agree’).

\begin{itemize}
    \item Self-efficacy: On the basis of the scale provided by Ajzen \cite{RN81}, the self-efficacy scale used in the current study was customized for water and energy conservation on university campuses. The items include (i) “\textit{I believe I have the ability to avoid unnecessary electricity and water use in the dormitory}”, (ii) “\textit{I give up when I feel it is difficult to implement water and energy conservation}”, and (iii) “\textit{How much control do you feel you have over the water and energy consumption in your dormitory? (1=Very little control, ..., 5=Complete control)}”. Item (ii) was reverse scored. The Cronbach's $\alpha$ of this scale was found to be 0.62.
    
    \item Output expectations: Following the practice of Thøgersen and Grønhøj \cite{RN44}, three types of output expectations were measured: expected physical outcomes of conservation efforts, expected social outcomes (or subjective norms), and expected self-evaluative outcomes (or personal norms). A total of 5 items were included: (i) “\textit{I can contribute to slowing down climate change and resource depletion through saving water and energy}” for expected physical outcomes; (ii) “\textit{My family expects me to save water and energy}”; and (iii) “\textit{My friends believe that I should make efforts to save water and energy}” for expected social outcomes; (iv) “\textit{It makes me feel like a better person to save water and energy in my dormitory}”; and (v) “\textit{I feel bad if I use more water and energy resources than necessary}” for expected self-evaluative outcomes. Item (v) is reverse scored. The Cronbach's $\alpha$ of this scale was found to be 0.83.
    
    \item Perceived impediments/facilitators: The items were designed following the scale used in Thøgersen and Grønhøj \cite{RN44}. The items include (i) “\textit{As the conditions are, it is difficult to avoid unnecessary water and energy consumption in my dormitory}” and (ii) “\textit{It takes a big extra effort from me to avoid unnecessary water and energy consumption in my dormitory.}” The Cronbach's $\alpha$ of this scale was found to be 0.80.
    
    \item Attitudes: The attitudes scale included 5 items that measured values, sense of responsibility, and interest in water and energy conservation. These items were designed in accordance with the Theory of Planned Behaviors and previous research regarding the impact of attitudes on behavior intentions: (i) “\textit{It’s important to conserve water and energy}” for value; (ii) “\textit{Resource depletion is irrelevant to me}”; and (iii) “\textit{As a student, I should take responsibility for conserving water and energy on campus}” for sense of responsibility; (iv) “\textit{I encourage others to conserve water and energy}”; and (v) “\textit{I enjoy learning about methods to conserve water and energy}” for interest. Item (ii) is reverse scored. The Cronbach's $\alpha$ of this scale was found to be 0.81.
    
    \item Neighborhood perceptions: The items were designed to measure respondents’ perceptions of the conservation behaviors of their neighborhoods. Two items from the measurement scale proposed by Thøgersen and Grønhøj \cite{RN44} were customized for the university campus context: (i) “\textit{My roommates conserve water and energy in their daily lives}” and (ii) “\textit{Apart from my roommates, people around me are very attentive to conserving water and energy in their daily lives}”. The Cronbach's $\alpha$ of this scale was found to be 0.77.
\end{itemize}

    \textbf{(3) Conservation intentions.} We measured conservation intentions through three main items: (i) “\textit{Read the delivered usage information}” for attention to the nudges, (ii) “\textit{Continuously pay attention to the water and energy usage in the dormitory}” for awareness change, and (iii) “\textit{Adjust water and energy usage habits to conserve}” for behavioral change. The participants were instructed to rate their agreement with each statement on a Likert scale ranging from 1 (‘strongly disagree’) to 5 (‘strongly agree’). As item (i) could be measured only in T1 and T2, it was presented in the Results section but was not included in the modeling section of this study. The Cronbach's $\alpha$ of this scale was found to be 0.74.

    \textbf{(4) Attitudes toward nudges.} The participants in T1 and T2 were requested to report their attitudes toward nudges by indicating their support for the nudges and rating the extent to which they believed the nudges helped them understand how to conserve more effectively and enabled them to conserve better. Additionally, the participants were asked to rank the effectiveness of each type of information, including usage amount, historical comparison, social comparison, conservation tips, and conservation potential.

    The six scales utilized for subsequent modeling (all cognition factors and conservation intentions) demonstrated acceptable composite reliabilities (Cronbach’s $\alpha$ values are reported alongside the items for each scale below) and validities (adequate fit of CFA models: $\chi^2$(df) = 5.78, RMSEA = 0.06, CFI = 0.94, SRMR = 0.04). Detailed information on the items is provided in Supplementary Table S1.

\subsection{Quantifying the increases in conservation intentions}

    In this study, the average conservation intention changes when receiving the nudges, i.e., the average treatment effect, were evaluated by comparing the average conservation intentions among C, T1, and T2. The individual-level changes in the conservation intentions of all the samples, i.e., individual treatment effects, were estimated through a multi-arm causal forest. The heterogeneity of nudges on different social groups was also analyzed on the basis of the results of the causal forest.

    We first determined the average changes by examining the differences in average conservation intentions among C, T1, and T2. The role of traditional nudges was assessed by comparing the differences between C and T1, and the further enhancements caused by LLM-powered nudges were highlighted through the comparison of T1 and T2. A two-step statistical test was conducted; we first employed the Kruskal‒Wallis test, which is a nonparametric alternative to analysis of variance (ANOVA), which tests the null hypothesis that the medians of all groups are equal. We then conducted Dunn’s test and Conover’s test to further ascertain pairwise differences. Significant differences, or improvements, in conservation intentions with more comprehensive nudge content should have been observed. After the statistical tests, we quantified the average conservation intention changes by calculating the differences in average conservation intentions among the groups.

    To further estimate the distribution of individual-level changes, the causal forest algorithm \cite{RN64} was used in this study. The algorithm can help acquire better estimations by minimizing the need for parametric assumptions and maximizing the out-of-sample predictive accuracy. The causal forests can be established by local centering (using nuisance models to remove confounding factors) and conservation intention change estimation (fitting a model to cluster observed heterogeneity in conservation intention changes, form an adaptive kernel, and then estimate the conservation intention changes within each kernel bandwidth).

    (1) Local centering. Two random forests were employed, namely, one $m^*$ that predicted the potential outcomes and one $e^*$ that predicted the propensity score (i.e., the possibility of treatment). Using these two models, we obtained the residuals for each observation as follows:

    \begin{equation}
        Y_i^c=Y_i-m^* (X_i)
    \end{equation}

    \begin{equation}
        W_i^c=W_i-e^* (X_i)
    \end{equation}

    This step essentially used the residuals from the models as the treatment and outcome variables in the final model, which partially eliminated selection effects. Even though this study employed a randomized controlled trial, this step was retained to ensure the rigor of the estimation results.

    (2) Estimating changes in conservation intention. A heterogeneity model, which is a random forest composed of causal trees, was established. It differs from ordinary random forests. First, a special splitting criterion, which is designed to uncover heterogeneity in conservation intention changes by estimating the expected squared error, is used. Second, it is “honest”, indicating that each tree grows on the basis of half of its sample and estimates the terminal node using the other half \cite{RN82}. This approach prevents overfitting and ensures that the predictions of each tree are asymptotically normal. The causal trees are collectively trained as an R-learner meta-learner \cite{RN83}, meaning that it minimizes the R-loss function to estimate the following heterogeneity model $\tilde{\tau}$:

    \begin{equation}
        \tau(\cdot)=\arg\min_{\tau}\left\{\frac{1}{n}\sum_{i=1}^{n}[Y_{i}^{c}-W_{i}^{c}\tau(X_{i})]^{2}\right\}+\Lambda_{n}\{\tau(\cdot)\}
    \end{equation}

    where $n$ is the sample size; $Y_{i}^{c}$ and $W_{i}^{c}$ are the centered values of the outcome and treatment, respectively; and $\Lambda_{n}\{\tau(\cdot)\}$ is a regularization term.

    In this study, the algorithm was implemented using the generalized random forest (\textit{grf}) R package. A total of 10,000 trees were grown. Socio-structural and cognitive factors collected in the questionnaire were used as features in the model training process. A multi-arm causal forest was established to quantify the conservation intention changes between C and T1 and between C and T2.
    
    The robustness of the multi-arm causal forest was validated through bootstrap resampling, placebo tests, heterogeneity tests, and calibration tests. The details and results of these validation methods are provided in Supplementary Notes S3-S6 and Supplementary Tables S3-S6.

\subsection{Explanation of conservation intentions and the function of LLMs}

    Existing evidence indicates that individual water and energy conservation behaviors are influenced by numerous psychological and social structural factors, particularly self-efficacy \cite{RN41,RN42}, perceived impediments in environmental conditions \cite{RN45,RN46}, and attitudes toward conservation \cite{RN49,RN50}. Self-efficacy and perceived environmental conditions (i.e., neighborhood perceptions) are key constructs in Bandura’s theory of social cognition \cite{RN68}, whereas the shaping of behavioral intentions by attitudes is the core viewpoint of Ajzen and Fishbein’s theory of reasoned action \cite{RN69}. Both theories have been widely used in previous studies aimed at explaining behavioral intentions \cite{RN84,RN85}. This makes these two theories potentially useful reference frameworks for the current study, allowing us to simultaneously consider the relationships among the aforementioned factors and their impact on conservation intentions. Since the theory of social cognition is comprehensive and complex and follows the approach of most previous studies that have employed these two theories, we designed a simplified version based on individual energy and water conservation to simulate the formation mechanism of conservation intentions.

    Specifically, the theory of reasoned action helps explain the relationship between attitudes and conservation intentions. Integrating these theories, our conceptual framework posits that the core forming individual behavior is self-efficacy, which is an individual’s level of confidence in executing a specific behavior. This is also presented as common sense in previous studies, such as Wang et al. \cite{RN86}. An individual’s output expectations and values attached to given outcomes (i.e., impediments) are influenced by self-efficacy, which in turn affects the generation of reasoned behavior decisions (i.e., attitudes influencing conservation intentions). In addition to these self-cognitive factors, the process of individuals observing and learning from others’ behaviors, referred to as neighborhood perceptions, is also emphasized, as it can shape all self-cognitive factors and conservation behaviors.

    A structural equation model, i.e., the \textit{full model}, was first established to examine the proposed framework of the decision-making process of conservation intentions. Keeping the same paths and structure as the \textit{full model}, the role of LLM-powered nudges was further explored and highlighted by establishing and comparing structural equation models for C, T1 and T2.

    To maintain statistical power due to reduced sample sizes in subgroup analyses, we shifted from using latent variables to a path model using variable means for the subgroup models. Before proceeding with the subgroup models, we verified the measurement invariance of the scales (except for conservation intention, as we expected to see differences among groups within our study) to ensure the validity of multigroup SEM analysis, as emphasized by Putnick and Bornstein \cite{RN87}. Detailed results are provided in Supplementary Note S2 and Supplementary Table S2.

    All the structural equation models were established through \textit{Mplus} 8.0 following the guidance of Muthén and Muthén \cite{RN88}. We evaluated the models via the comparative fit index (CFI), root mean square error of approximation (RMSEA) and standardized root mean square residual (SRMR). A CFI value close to or above 0.90, an RMSEA value close to or below 0.08, and an SRMR value close to or smaller than 0.08 were considered acceptable \cite{RN89}.

\section*{Acknowledgments}
    This study was supported by National Natural Science Foundation of China (grant no. 52470212), and Self-Funded Project of Institute for Global Industry, Tsinghua University (grant no. 202-296-001, 2024-06-18-LXHT002).

\section*{Ethics declarations}
\subsection*{Competing interests}
    The authors declare no competing interests.

\section*{Supplementary Information}
    Supplementary Figure S1, Supplementary Notes S1-S7, and Supplementary Tables S1-S7.

%----------------------------------------------------------
\printbibliography

%----------------------------------------------------------

\clearpage


\section*{Supplementary material}

# Supplementary Information

## Contents

### Validation of the randomness of the grouping

### Validation of the scales

### Validation of the multi-arm causal forest

### Details of the structural equation models

### Calculation of potential environmental benefits

## Validation of the randomness of the grouping

**Supplementary Note S1**    Explanations on randomness of the grouping.

As described in the Methods section, each participant was randomly assigned to a group upon entering the questionnaire survey. We examined the differences in levels of cognition and social-structural factors used in the causal forest and structural equation models (as illustrated in Supplementary Figure S1) among the three groups using the same two-step statistical testing procedure employed to identify differences in conservation intentions. First, we conducted the Kruskal-Wallis test, a non-parametric alternative to analysis of variance (ANOVA), followed by Dunn's Test and Conover's Test to further assess pairwise differences.

We observed no significant differences in these tests, indicating that there were no significant differences among the three groups.

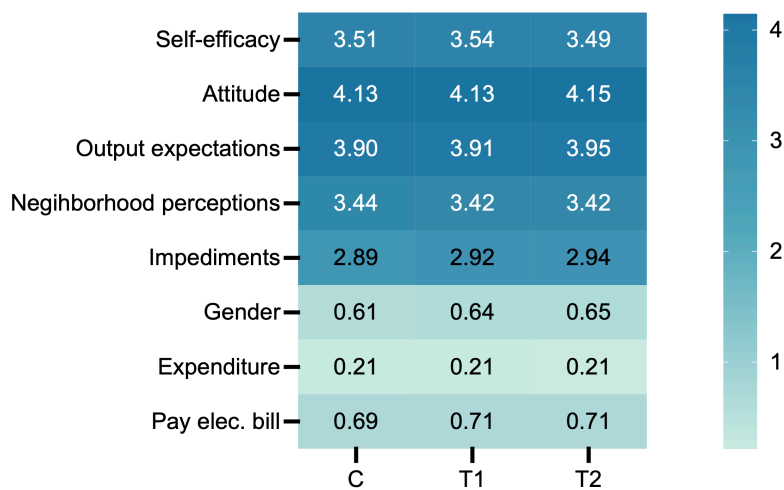

**Supplementary Figure S1**    Levels of cognition and social-structural factors among three groups.

## Validation of the scales

### (1) Confirmatory factor analysis (CFA) results.

**Supplementary Table S1   CFA results**

| Factor | | Item | Estimate | SE | 95% CI Lower | 95% CI Upper | p |
|---|---|---|---|---|---|---|---|
| Self-efficacy | (i) | I believe I have the ability to avoid unnecessary water and energy use in the dormitory. | 0.63*** | 0.02 | 0.59 | 0.68 | <.0001 |
| | (ii) | I give up when I feel it is difficult to implement water and energy conservation. | 0.49*** | 0.02 | 0.44 | 0.53 | <.0001 |
| | (iii) | How much control do you feel you have over the water and energy consumption in your dormitory? (1=Very little control, … ,5=Complete control) | 0.44*** | 0.03 | 0.39 | 0.50 | <.0001 |
| Output expectations | (i) | I can contribute to slowing down climate change and resource depletion through saving water and energy. | 0.73*** | 0.01 | 0.70 | 0.76 | <.0001 |
| | (ii) | My family expects me to save water and energy. | 0.79*** | 0.01 | 0.77 | 0.81 | <.0001 |
| | (iii) | My friends believe that I should make efforts to save water and energy. | 0.67*** | 0.02 | 0.64 | 0.70 | <.0001 |
| | (iv) | It makes me feel like a better person to save water and energy in my dormitory. | 0.64*** | 0.02 | 0.61 | 0.67 | <.0001 |
| | (v) | I feel bad if I use more water and energy resources than necessary. | 0.66*** | 0.02 | 0.63 | 0.69 | <.0001 |
| Perceived impediments | (i) | As the conditions are, it is difficult to avoid unnecessary water and energy consumption in my dormitory. | 0.81*** | 0.02 | 0.77 | 0.85 | <.0001 |
| | (ii) | It takes a big extra effort from me to avoid unnecessary water and energy consumption in my dormitory. | 0.82*** | 0.02 | 0.77 | 0.86 | <.0001 |
| Attitude | (i) | It's important to conserve water and energy. | 0.66*** | 0.02 | 0.62 | 0.69 | <.0001 |
| | (ii) | Resource depletion is irrelevant to me. | 0.53*** | 0.02 | 0.49 | 0.57 | <.0001 |
| | (iii) | As a student, I should take responsibility for conserving water and energy on campus. | 0.77*** | 0.01 | 0.75 | 0.80 | <.0001 |
| | (iv) | I encourage others to conserve water and energy. | 0.72*** | 0.02 | 0.69 | 0.75 | <.0001 |
| | (v) | I enjoy learning about methods to conserve water and energy. | 0.71*** | 0.02 | 0.68 | 0.74 | <.0001 |
| Neighbor. perception | (i) | My roommates conserve water and energy in their daily lives. | 0.81*** | 0.02 | 0.78 | 0.85 | <.0001 |
| | (ii) | Apart from my roommates, people around me are very attentive to conserving water and energy in their daily lives. | 0.76*** | 0.02 | 0.72 | 0.80 | <.0001 |
| Conser. intent. | (i) | Continuously pay attention to the water and energy usage in the dormitory. | 0.88*** | 0.02 | 0.84 | 0.91 | <.0001 |
| | (ii) | Adjust water and energy usage habits to conserve. | 0.67*** | 0.02 | 0.63 | 0.70 | <.0001 |

### (2) Measurement invariance confirmation.

**Supplementary Note S2    Explanations on the measurement invariance.**

Measurement invariance assesses the psychometric equivalence of constructs across groups or over time and serves as the foundation for comparing results across groups[1]. We verified measurement invariance by examining changes in fit indices across different models.

Three models were constructed: M1—Configural Model, with loadings and intercepts freely estimated; M2—Metric Model, with loadings constrained to be equal; and M3—Scalar Model, with both loadings and intercepts constrained to be equal. Three types of

measurement invariance were tested: (1) **Configural invariance**, which assesses whether the factor structure is consistent across groups, typically determined by examining the fit of multiple-group CFA models; (2) **Metric invariance**, which evaluates whether the factor structure and item loadings are consistent across groups, typically determined by comparing differences between the Configural model and the Metric model—if established, path coefficients can be compared; (3) **Scalar invariance**, which examines whether the factor structure, item loadings, and item intercepts are consistent across groups, determined by comparing differences between the Configural model and the Scalar model—if established, latent means can be compared. For metric and scalar invariance, when the compared models fit well and the changes in RMSEA (ΔRMSEA), CFI (ΔCFI), and SRMR (ΔSRMR) are all ≤ 0.01, the corresponding invariance is confirmed[2].

The fit indices of the multiple-group CFA models are shown in Supplementary Table S2. The RMSEA, CFI, and SRMR for each model indicated good fit, and the differences in fit indices between M2 and M1, and between M3 and M1 (ΔRMSEA, ΔCFI, ΔSRMR), were all ≤ 0.01, confirming all three types of invariances.

**Supplementary Table S2**   Measurement invariance results of the scales.

| Model type | RMSEA | CFI | SRMR | Model comp | ΔRMSEA | ΔCFI | ΔSRMR |
|---|---|---|---|---|---|---|---|
| M1: Configural | 0.067 | 0.923 | 0.047 | - | - | - | - |
| M2: Metric | 0.065 | 0.923 | 0.052 | M1 | 0.002 | 0.000 | 0.005 |
| M3: Scalar | 0.066 | 0.917 | 0.054 | M1 | 0.001 | 0.006 | 0.007 |

## Validation of the multi-arm causal forest

### (1) Estimate through Bootstrap resampling

**Supplementary Note S3**   Explanations on the Bootstrap resampling method.

This study employed bootstrap resampling, performing 10,000 iterations, and recorded the average estimated changes in conservation intention for each resample. The results are presented in Supplementary Table S3. The 95% confidence intervals (CI) for the estimates of T1 and T2 are (0.08, 0.24) and (0.19, 0.34), respectively, indicating that the treatment effects are significant and non-zero.

**Supplementary Table S3**   Bootstrap resampling results of the multi-arm causal forest.

|  | Mean conservation intention changes | Std. Err. of conservation intention changes | Lower 95% CI | Upper 95% CI |
|---|---|---|---|---|
| C vs T1 | 0.16 | 0.04 | 0.08 | 0.24 |
| C vs T2 | 0.27 | 0.04 | 0.19 | 0.34 |

### (2) The placebo test

**Supplementary Note S4**   Explanations on the placebo test.

Placebo tests are essential tools that check whether the assumptions behind the research design are reasonable by examining what happens when those assumptions are slightly altered[3]. In the placebo test, we recalculated the average conservation intention changes through the same multi-arm causal forest used in the main text for 10,000 times, with each time the group assignment of participants was mixed and re-assigned randomly. We expected that, the average conservation intention changes in the placebo test would be close to zero and significantly different from the results based on the original groupings.

The results in Supplementary Table S4 show that the show that the average conservation intention changes for both placebo tests (C vs. T1 and C vs. T2) were -0.0001 and -0.0032, respectively. These estimates were significantly different from the original estimations (both p-values < 0.01), indicating that the random assignment did not produce meaningful treatment effects. This supports the validity of our original assumptions.

**Supplementary Table S4**   Results of the placebo test.

|  | Average conservation intention changes | | p value |
|---|---|---|---|
|  | Original | Placebo test |  |
| C vs T1 | 0.16 | -0.0001 | <0.01 |
| C vs T2 | 0.27 | -0.0032 | <0.01 |

### (3) The heterogeneity test

**Supplementary Note S5　　Explanations on the heterogeneity test.**

　　　To assess treatment effect heterogeneity, observations were divided into high and low conservation intention change groups based on whether their individually predicted changes exceeded the median. The average conservation intention change was then estimated separately for each subgroup using a double robust approach. By comparing the average changes between the high and low groups and calculating the 95% CI for their difference, we determined the presence and magnitude of heterogeneity in conservation intention changes.

　　　The results, shown in Supplementary Table S5, indicate that in both treatment comparisons (C vs T1 and C vs T2), the average conservation intention change for the high effect group is substantially higher than that for the low effect group. Additionally, the 95% CI for the changes do not include zero. This suggests that conservation intention changes vary meaningfully between subgroups and that the multi-arm causal forest effectively captured the variation among individuals.

**Supplementary Table S5　　Heterogeneity test results of the multi-arm causal forest.**

|  | Estimation of High effect group | Estimation of Low effect group | Differences (High – Low) | Difference Lower 95% CI | Difference Upper 95% CI |
|---|---|---|---|---|---|
| C vs T1 | 0.28 | 0.04 | 0.24 | 0.235 | 0.251 |
| C vs T2 | 0.41 | 0.13 | 0.28 | 0.271 | 0.290 |

### (4) The calibration test

**Supplementary Note S6　　Explanations on the calibration test.**

　　　The calibration test assesses the accuracy and reliability of a causal forest by performing a linear regression of the conservation intention changes on two predictors: the mean forest prediction and the differential forest prediction. Using held-out data, the test evaluates whether the average predictions are unbiased (indicated by a coefficient of ~1 for the mean forest prediction) and whether the model accurately captures heterogeneity in the data (indicated by a coefficient of ~1 for the differential forest prediction). Additionally, the p-value associated with the differential forest prediction coefficient serves as an omnibus test for the presence of heterogeneity. A significant p-value suggests that the model effectively captures the variability in the predictions.

　　　The results are shown in Supplementary Table S6. All the estimated coefficients are close to 1, demonstrating that the model's average predictions are accurate and that it effectively captures significant heterogeneity in the data. Also, the p-value of the differential forest prediction is highly significant, revealing a good capture of variability among individuals.

**Supplementary Table S6　　Calibration test results of the multi-arm causal forest.**

|  | Mean forest prediction | | | | Differential forest prediction | | | |
|---|---|---|---|---|---|---|---|---|
|  | Estimate | Std. Err. | t value | p value | Estimate | Std. Err. | t value | p value |
| C vs T1 | 1.01*** | 0.28 | 3.58 | <0.01 | 0.96*** | 0.36 | 2.64 | <0.01 |
| C vs T2 | 1.00*** | 0.16 | 6.14 | <0.01 | 1.19*** | 0.27 | 4.35 | <0.01 |

### Details of the structural equation models

**Supplementary Table S7　　The measurement part of the *full model*.**

| Path | Estimate | SE | 95% CI Lower | 95% CI Upper | P value |
|---|---|---|---|---|---|
| **MEASUREMENT PART: Self-efficacy** | | | | | |
| ITEM (i) | 0.70*** | 0.02 | 0.66 | 0.75 | <0.01 |
| ITEM (ii) | 0.45*** | 0.03 | 0.40 | 0.51 | <0.01 |
| ITEM (iii) | 0.51*** | 0.03 | 0.46 | 0.56 | <0.01 |
| **MEASUREMENT PART: Output expectations** | | | | | |
| ITEM (i) | 0.72*** | 0.02 | 0.69 | 0.74 | <0.01 |
| ITEM (ii) | 0.78*** | 0.01 | 0.76 | 0.81 | <0.01 |
| ITEM (iii) | 0.67*** | 0.02 | 0.63 | 0.70 | <0.01 |
| ITEM (iv) | 0.67*** | 0.02 | 0.64 | 0.70 | <0.01 |
| ITEM (v) | 0.69*** | 0.02 | 0.66 | 0.72 | <0.01 |

| Path | Estimate | SE | 95% CI Lower | 95% CI Upper | P value |
|---|---|---|---|---|---|
| **MEASUREMENT PART: Perceived impediments** | | | | | |
| ITEM (i) | 0.82*** | 0.03 | 0.77 | 0.88 | <0.01 |
| ITEM (ii) | 0.80*** | 0.03 | 0.75 | 0.86 | <0.01 |
| **MEASUREMENT PART: Attitude** | | | | | |
| ITEM (i) | 0.64*** | 0.02 | 0.61 | 0.68 | <0.01 |
| ITEM (ii) | 0.51*** | 0.02 | 0.47 | 0.55 | <0.01 |
| ITEM (iii) | 0.76*** | 0.01 | 0.73 | 0.79 | <0.01 |
| ITEM (iv) | 0.75*** | 0.01 | 0.72 | 0.77 | <0.01 |
| ITEM (v) | 0.73*** | 0.01 | 0.71 | 0.76 | <0.01 |
| **MEASUREMENT PART: Neighborhood perceptions** | | | | | |
| ITEM (i) | 0.79*** | 0.02 | 0.75 | 0.82 | <0.01 |
| ITEM (ii) | 0.76*** | 0.02 | 0.73 | 0.80 | <0.01 |
| **MEASUREMENT PART: Conservation intentions** | | | | | |
| ITEM (i) | 0.85*** | 0.02 | 0.81 | 0.89 | <0.01 |
| ITEM (ii) | 0.68*** | 0.02 | 0.64 | 0.72 | <0.01 |

## Calculation of potential environmental benefits

**Supplementary Note S7**  Explanations on the calculation of potential environmental benefits mentioned in the Discussion section.

First, the total electricity and water consumption in student dormitories are calculated based on the per capita usage for different academic levels (undergraduates, master's students, doctoral students) and the number of enrolled students. Then, the total water consumption is multiplied by 8.3% (the average conservation intention change calculated in the main text, as revealed in **Figure 3a**) to determine the potential savings. Results show that there would potentially be electricity savings of 3.36 million kWh per year and water savings of 0.11 million cubic meters per year if students in the case university could fully implement these conservation behaviors under LLM-powered nudges.

According to the *Statistical report on China's educational achievements in 2023*[4], there are a total of 20.35 million enrolled undergraduate students, 3.27 million enrolled master's students, and 0.61 million enrolled doctoral students nationwide. Due to the lack of comprehensive national statistics on total water consumption or per capita usage in higher education institutions, data from the case campus is utilized for estimation. The nationwide total water consumption in student dormitories is estimated by multiplying the per capita usage for each academic level by the respective number of enrolled students. Then, the total consumption is multiplied by 8.3% to obtain the potential savings. Results show that if all universities in China use LLM-powered nudges, there would be electricity savings of 1.14 billion kWh per year and water savings of 0.04 billion cubic meters per year.

In 2023, the per capita residential water usage in China was 125 liters per day[5], and the per capita residential electricity usage was approximately 2.62 kWh per day (calculated based on the total residential electricity consumption of 1,352.4 billion kWh divided by the total population of 1.412 billion[6]). Consequently, the potential electricity savings could support approximately 1.20 million people for one year, and the potential water savings could support approximately 863,796 people for one year.

## Supplementary References



\end{document}